\documentclass[aps,superscriptaddress, twocolumn,showpacs, bibliography]{revtex4-1}
\pdfoutput=1
\usepackage{color}
\usepackage{amsmath, amsthm, amsfonts}    
\usepackage{amssymb}
\usepackage{mathtools}
\usepackage{framed} 
\usepackage{mathptmx} 
\usepackage[table]{xcolor}
\usepackage{dsfont}
\usepackage{enumitem} 
\usepackage[]{mdframed}
\usepackage{appendix}
\usepackage{braket}
\usepackage[skins,theorems]{tcolorbox}
\tcbset{highlight math style={enhanced,
  colframe=red,colback=white,arc=0pt,boxrule=1pt}}
\usepackage{cancel}
\usepackage{empheq}
\usepackage{lipsum}
\usepackage{graphicx}
\usepackage{tikz}
\usepackage{mdframed}
\usepackage{bigints}
\usepackage{makecell}
\usepackage[unicode=true,pdfusetitle,
 bookmarks=true,bookmarksnumbered=false,bookmarksopen=false,
 breaklinks=true,pdfborder={0 0 0},backref=false,colorlinks=true,citecolor=blue]{hyperref}
\usepackage{graphicx}   
\usepackage[caption=false]{subfig}       
\usepackage{bm}            
\usepackage[normalem]{ulem}  

\newlength\imagewidth
\newlength\imagescale
\def\be{\begin{eqnarray}}
\def\ee{\end{eqnarray}}
\def\r{{\bf r}}

\def\E{{\bf E}}
\usepackage{tabstackengine}
\setstackEOL{\cr}

\AtBeginEnvironment{pmatrix}{\setlength{\arraycolsep}{1pt}}

\DeclareUnicodeCharacter{2212}{-}

\definecolor{JOT-color}{named}{blue}
\definecolor{CSF-color}{named}{orange}

\begin{document}

\title{The Stokes Vector Measurement}

\author{Jorge Olmos-Trigo}
\email{jolmostrigo@gmail.com}
\affiliation{Departamento de Física, Universidad de La Laguna, Apdo. 456. E-38200, San Cristóbal de La Laguna, Santa Cruz de Tenerife, Spain.}

\begin{abstract}
The multipolar expansion of the electromagnetic field plays a key role in the study of light-matter interactions.  All the information about the radiation and coupling between the incident wavefield and the object is embodied in the electric and magnetic scattering coefficients $\{a_{\ell m}, b_{\ell m} \}$ of the expansion. However, the experimental determination of $\{a_{\ell m}, b_{\ell m} \}$ requires measuring the components of the scattered field in all directions, something that is exceptionally challenging. Here, we demonstrate that a single measurement of the Stokes vector unlocks access to the quadrivector $ \mathbf{D}_{\ell m}  = \left[|a_{\ell m}|^2, |b_{\ell m}|^2, \Re \{ a_{\ell m} b^*_{\ell m}  \}, \Im \{ a_{\ell m} b^*_{\ell m}  \} \right]$. Thus, our Stokes polarimetry method allows us to capture $|a_{\ell m}|^2$ and $|b_{\ell m}|^2$ separately, a distinction that can not be achieved by measuring the total energy of the scattered field via an integrating sphere. Importantly, we demonstrate the robustness of our Stokes polarimetry method, showing its fidelity with just two measurements of the Stokes vector at different scattering angles.  Our findings, supported by analytical theory and exact numerical simulations, can find applications in Nanophotonics and greatly facilitate routine light-scattering measurements in optical laboratories. 
\end{abstract}

\maketitle

{\emph{Introduction.}}--- The multipolar expansion of the electromagnetic field is a key tool in the study of light-matter interactions and has historically played a pivotal role in several branches of Nanophotonics~\cite{devaney1974multipole}. These include optical forces~\cite{barton1989theoretical}, optical torques~\cite{marston1984radiation}, and chiral light-matter interactions~\cite{bohren1974light},  among others~\cite{fernandez2015exact, alaee2018electromagnetic}.
The multipolar expansion of the electromagnetic field is typically written as an infinite sum of electric and magnetic vector spherical harmonics that are, in turn, weighted by its corresponding electric and magnetic scattering coefficients, respectively~\cite{jackson1999electrodynamics}. 
Researchers  have access to the multipolar expansion of the incident wavefield since its coefficients are known. However, the situation changes when the incident wavefield interacts with an object. In this case, the electric and magnetic scattering coefficients, denoted as $a_{\ell m}$ and $b_{\ell m}$, respectively, are unknown complex quantities and their determination is crucial to solving the scattering problem under investigation. In this setting,  $\ell$ and $m$ denote the multipolar order and total angular momentum, respectively~\cite{jackson1999electrodynamics}.

\begin{figure*}[t!]
\centering
\includegraphics[width=\textwidth]{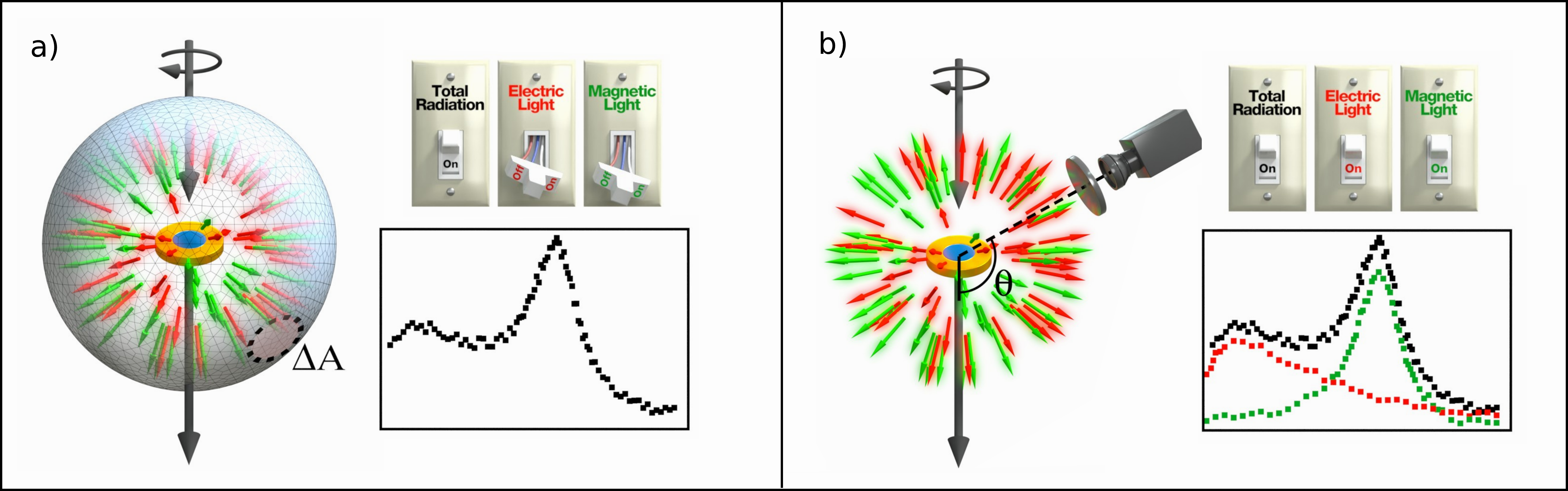}
\caption{Artistic representation of the measurement of the scattering cross-section of a nanodisk under the illumination of a circularly polarized wavefield. The red and green arrows represent the electric and magnetic amplitudes of the scattering coefficients in the scattered field.  (a) An integrating sphere is placed in the far-field to collect the components of the scattered field in all directions. This measurement does not allow distinguishing between the electric and magnetic amplitudes of the scattering coefficients. (b) The Stokes vector measurement, in which a photo-diode and conventional waveplates are placed at an scattering angle $\theta$, allows telling between the electric and magnetic contributions to the scattering cross-section.}
\label{Fig_1}
\end{figure*}

From the theoretical perspective, the following multi-step procedure is used to retrieve $\{a_{\ell m}, b_{\ell m} \}$: First, numerical methods are employed to obtain the components of the scattered field in all directions~\footnote{When the object has spherical symmetry, Mie theory can be directly employed.}. Some examples of these numerical approaches are the T-matrix method~\cite{mackowski1996calculation}, the Discrete Dipole Approximation (DDA)~\cite{draine1994discrete}, along with all kinds of Maxwell solvers.
Subsequently, by  projecting the scattered field onto the corresponding electric (or magnetic) vector spherical harmonic,  the desired electric (or magnetic) scattering coefficient can be calculated~\cite{jackson1999electrodynamics}.
However, a fundamental problem arises in the previous approach to determine $\{a_{\ell m}, b_{\ell m} \}$: it lacks experimental equivalence, primarily due to the formidable task of measuring the components of the scattered field in all directions. 
Here, we present a Stokes polarimetry approach that solves this experimental challenge for objects well-described by a single multipolar order $\ell$ and total angular momentum $m$.
More specifically, we demonstrate that a measurement of the Stokes vector grants access to all  the components of the quadrivector 
$\mathbf{D}_{\ell m} = \left[|a_{\ell m}|^2, |b_{\ell m}|^2, \Re \{ a_{\ell m} b^*_{\ell m}  \}, \Im \{ a_{\ell m} b^*_{\ell m}  \} \right]$,  enabling the separate detection of $|a_{\ell m}|^2$ and $|b_{\ell m}|^2$. 
Remarkably, this distinction between the electric and magnetic amplitudes of the scattering coefficients is unreachable if measuring the scattering cross-section.
To visualize this distinction, check Fig.~\ref{Fig_1}, where we show the scattering cross-section of a nanodisk excited by a circularly polarized wavefield. Two experimental setups are depicted to measure the scattering cross-section:  an integrating sphere embedding the excited nanodisk (see Fig.~\ref{Fig_1}a) and our Stokes polarimetry approach in which only a photo-diode and conventional wave-plates are needed  (see Fig~\ref{Fig_1}b).  As Fig~\ref{Fig_1}b shows, the Stokes polarimetry approach allows telling between $|a_{\ell m}|^2$ and $|b_{\ell m}|^2$.

Importantly, in our work, we do not impose any restrictions on the incident wavefield. Accordingly, our Stokes polarimetry approach can accommodate a wide range of illumination conditions. On top of that, by measuring the Stokes vector at two different scattering angles, we establish the fidelity of our Stokes polarimetry approach. That is, we demonstrate that our method is experimentally robust and can be trusted without performing any numerical simulation. 
Consequently, our findings, supported by analytical theory and exact numerical simulations, are promising for all branches of Nanophotonics as they greatly facilitate the experimental characterization of objects in optical laboratories. 

{\emph{The Stokes vector and the multipolar expansion of the field.}}--- The Stokes vector $\mathbf{S} = [s_0, s_1, s_2,  s_3]$  {unambiguously} describe the polarization state and energy flux of any electromagnetic field in the far-field limit~\cite{stokes1851composition}. Importantly, the components of the Stokes vector, typically referred to as the Stokes parameters~\cite{crichton2000measurable}, can be measured using a photo-diode and conventional wave-plates~\cite{hinamoto2020colloidal, negoro2023helicity}. Following Bohren's and Huffman book~\cite{bohren2008absorption}, the Stokes parameters read as
\begin{align} \label{s_0}
s_0 &= |E_\theta|^2 + |E_\varphi|^2, 
\\
s_1 &= |E_\theta|^2 - |E_\varphi|^2, 
\\
s_2 &=  -2\Re \{E_\theta E^*_\varphi \}, \\\label{s_3}
 s_3 &= 2\Im \{E_\theta E^*_\varphi \}. 
\end{align}
Here $\Re$ and $\Im$ denote the real and imaginary parts, respectively. 
By inspecting Eqs.~\eqref{s_0}-\eqref{s_3}, we note that $s_0$ is the total scattered intensity, $s_1$ is the degree of linear polarization, $s_2$ is the degree of linear polarization at 45 degrees, and $s_3$ denotes the degree of circular polarization~\cite{bohren2008absorption}.
To determine the Stokes parameters, we first need to obtain the transversal components of the scattered field, namely, $E_\theta$ and $E_\varphi$, evaluated in the radiation (far) zone. 
Hereafter, we follow Jackson's notation in its third edition to describe the multipolar expansion of the scattered field~\cite{jackson1999electrodynamics}.
After some algebra (see Supporting Information S1 for the detailed calculation), it can be shown that the scattered  field $\E (k \r)$ can be  written in the radiation zone (when $ kr \rightarrow \infty$) as  
\begin{equation} \label{E_far_1}
\lim_{kr \rightarrow \infty} \E (k \r) = \left[ E_\theta \hat{\mathbf{e}}_{{{\theta}}} + E_\varphi \hat{\mathbf{e}}_{{{\varphi}}}\right], 
\end{equation}
where 
\be \label{key_1}
E_\theta &=& {E_0} \sum_{\ell m} \bar{C}_{\ell  m} (kr, \varphi) \left[  a_{\ell m} \tau_{\ell m} ({\theta}) - im b_{\ell m} \pi_{\ell m} ({\theta}) \right],  \\ \label{key_2}
E_\varphi &=& {E_0} \sum_{\ell m}  \bar{C}_{\ell  m} (kr, \varphi) \left[ im  a_{\ell m} \pi_{\ell m} ({\theta}) + b_{\ell m} \tau_{\ell m} ({\theta}) \right].
\ee
Here $E_0$ is the amplitude of the incident wavefield,  $k$ is the radiation wavenumber,  $r = |\r|$ denotes the observation distance to the center of the object, and  $\theta$ and $\varphi$ denote the scattering and azimuthal angles, respectively. Moreover, we have defined~\footnote{Interestingly, $\pi_{lm}(\theta)$ and $\tau_{l m} (\theta)$  are real-valued functions that Bohren and Huffman defined to tackle the absorption and scattering by a sphere for $m = 1$  (see Eq. 4.46 of Ref.~\cite{bohren2008absorption}).} 
\begin{align} \label{chi_tau}
\pi_{\ell m}(\theta) =  \frac{P^{m}_{\ell }(\cos \theta)}{\sin \theta}, &&    \tau_{\ell m}(\theta)  = \frac{d P^{m}_{\ell }(\cos \theta)}{d{\theta}},
\end{align}
where $P^{m}_{\ell}(\cos \theta)$ are the Associated Legendre Polynomials~\cite{jackson1999electrodynamics} and
\begin{equation} \label{constant}
\bar{C}_{\ell  m} (kr, \varphi)= \frac{e^{ikr}}{kr} \left[ \frac{(-i)^{\ell + 2}}{\sqrt{\ell ( \ell + 1) }} \sqrt{\frac{2 \ell +1}{4\pi}\frac{(\ell - m)!}{(\ell+m)!}} \right]e^{im \varphi}.
\end{equation}

\begin{table*}[]
    \caption{Receipt to use the Stokes vector measurement to experimentally characterize objects.}
    \setlength{\fboxrule}{1pt} 
    \fbox{
        \parbox{0.975\textwidth}{
            \begin{enumerate}
                \item Measure the Stokes vector $\mathbf{S}$ at an angle  $\theta_1$. Note that this experimental measurement takes into account all multipoles.

                \item Calculate the matrix $U_{\ell m}$ for fixed values of $\ell$ and $m$ at $\theta_1$. An example of the calculation of $U_{\ell m}$, for instance, $U_{11}$, can be found in the Supporting Information S2.

                \item Use Eq.~\eqref{compact} to obtain $\mathbf{D}_{\ell m}$. 

                \item Repeat steps 1, 2, and 3 for a different scattering angle $\theta_2$.

                \item Compare the values of $\mathbf{D}_{\ell m}$ evaluated at $\theta_1$ and $\theta_2$:
                
                \begin{enumerate}
                    \item If they resemble each other, the scattering can be fully described by a single multipolar order $\ell$ and $m$, and no additional measurement is needed. In this scenario, $\mathbf{D}_{\ell m}$ is the correct quadrivector: the object has been successfully characterized.

                    \item If they are different, then the excited object cannot be described by the selected values of $\ell$ and $m$. 
                \end{enumerate}   
            \end{enumerate}
        }
    }
    \label{T_1}
\end{table*}

{\emph{The electric and magnetic scattering coefficients from the Stokes vector.}}---At this point, we have all the ingredients to calculate the Stokes vector $\mathbf{S}$. 
To that end, let us insert Eqs.~\eqref{key_1}-\eqref{key_2} into Eqs.~\eqref{s_0}-\eqref{s_3} assuming that the object can be fully described by a single multipolar order $\ell$ and  total angular momentum $m$. After some algebra, it can be shown that
\begin{align} \label{s0_new}
\tilde{s}_0 &=  (|a_{\ell m}|^2 + |b_{\ell m}|^2)\gamma_{\ell m}(\theta)  - 4 \Im \{a_{\ell m} b^*_{\ell m} \} \eta_{\ell m}(\theta), \\
\tilde{s}_1 &= (|a_{\ell m}|^2 - |b_{\ell m}|^2)\nu_{\ell m}(\theta), \\ \label{mystery}
\tilde{s}_2 &= -2 \Re \{a_{\ell m} b^*_{\ell m} \} \nu_{\ell m}(\theta), \\ \label{s3_new}
\tilde{s}_3 &= 2 \left[\Im \{a_{\ell m} b^*_{\ell m} \}\gamma_{\ell m}(\theta) -(|a_{\ell m}|^2 + |b_{\ell m}|^2)\eta_{\ell m}(\theta) \right].
\end{align}
Here, we have defined  ${\mathbf{S}} = |E_0|^2|\bar{C}_{\ell m}(kr, \varphi)|^2 \mathbf{\tilde{S}}$ along with
\begin{align} \label{def_1}
 \gamma_{\ell m}(\theta) &= \left[ \tau_{\ell m} ^2(\theta) + m^2 \pi_{\ell m} ^2(\theta)\right], \\
 \eta_{\ell m}(\theta)  &= m   \tau_{\ell m} (\theta)\pi_{\ell m} (\theta), \\ \label{def_2}
 \nu_{\ell m}(\theta) &= \left[ \tau_{\ell m} ^2(\theta) - m^2 \pi_{\ell m} ^2(\theta)\right]. 
\end{align}
Let us briefly discuss the underlying physics behind Eqs.~\eqref{s0_new}-\eqref{s3_new}. These relations give the dimensionless Stokes vector $\tilde{\mathbf{S}}$ as a function of quadratic combinations of the electric and magnetic scattering coefficients of the multipolar expansion. Note that $ \gamma_{\ell m}(\theta)$, $ \eta_{\ell m}(\theta)$, and  $\nu_{\ell m}(\theta)$ do not depend on the optical response of the object and can be straightforwardly determined from Eq.~\eqref{chi_tau}. Now, from  Eqs.~\eqref{s0_new}-\eqref{s3_new} it is clear that if $a_{\ell m}$ and $b_{\ell m}$ are known, then the Stokes parameters can be calculated.
However, in an experiment, one does not have access to $a_{\ell m}$ and $b_{\ell m}$. In contrast, and as previously mentioned, the Stokes vector ${\mathbf{S}}$ can be measured using a photo-diode and conventional wave plates. 
Therefore, it is convenient to express $a_{\ell m}$ and $b_{\ell m}$ in terms of  ${\mathbf{S}}$.
In this regard, it is of utmost importance to note that by simply measuring the Stokes parameters, one cannot distinguish between the electric and magnetic amplitudes of the scattering coefficients. A key step remains to be done in order to achieve this important distinction~\footnote{Hereafter,  the $\theta$, $\varphi$, and $kr$ dependence will be assumed.}. 

Taking all the previous information into account, we can rewrite 
Eqs.~\eqref{s0_new}-\eqref{s3_new} as
\begin{equation} \label{compact}
\mathbf{D}_{\ell m} =   U_{\ell m}  {{\mathbf{S}}},
\end{equation}
\begin{align} \label{matrix_change}
U_{\ell m} = \frac{1}{A_{\ell m}}   
    \begin{pmatrix}
    \gamma_{\ell m} & \nu_{\ell m}& 0 & 2\eta_{\ell m}\\
    \gamma_{\ell m}& -\nu_{\ell m}& 0 & 2\eta_{\ell m} \\
       0 & 0 & -\nu_{\ell m} & 0\\
    2\eta_{\ell m} & 0 & 0 &  \gamma_{\ell m}
    \end{pmatrix},
\end{align}
with $A_{\ell m} = 2|E_0|^2|\bar{C}_{\ell m}|^2 \nu^2_{\ell m}$, and 
\begin{align} \label{vector}
\mathbf{D}_{\ell m} = 
 \begin{pmatrix}
    |a_{\ell m}|^2\\
    |b_{\ell m}|^2 \\ \Re \{a_{\ell m} b^*_{\ell m} \}\\
    \Im \{a_{\ell m} b^*_{\ell m} \}
    \end{pmatrix}, &&
    {\mathbf{{S}}} = 
 \begin{pmatrix}
    {s}_0\\
    {s}_1\\ {s}_2\\
    {s}_3
    \end{pmatrix}.
\end{align}
Equations~\eqref{compact}-\eqref{vector} are important results of this work: the quadrivector $\mathbf{D}_{\ell m}$, which dictates the radiation and coupling between the incident wavefield and the object, can be calculated by measuring the Stokes vector. 
Importantly, we have not made any assumption on the nature of the incident wavefield. Thereby, equations~\eqref{compact}-\eqref{vector} can be applied under general illumination conditions: a typical plane wave but also twisted (structured) light such as Gaussian and Laguerre-Gaussian beams with well-defined angular momentum of light~\cite{zambrana2012excitation, das2015beam}. Moreover, Eqs~\eqref{compact}-\eqref{vector} introduce an unprecedented advantage~\cite{lasaalonso2023characterizing}: the capacity to distinguish the electric and magnetic amplitudes of the scattering coefficients.
For a clearer understanding,  in Table~\ref{T_1} we present the steps to use and implement our Stokes-polarimetry method. 


{\emph{The physical properties of  $\mathbf{D}_{\ell m}$.}} To get a deeper insight into the relevance of our findings, we now discuss the features of each of the components that conform $\mathbf{D}_{\ell m}$. 

\textbullet \; $|a_{\ell m}|^2$ and $|b_{\ell m}|^2$: These scalar terms give full access to the electric and magnetic contribution to the scattering cross-section $\sigma_{\rm{sca}}$~\cite{jackson1999electrodynamics}. To show this fact, let us derive the scattering cross-section using the standard equation~\cite{jackson1999electrodynamics}
\begin{equation} \label{sigma_sca}
k^2 \sigma_{\rm{sca}} = \int_{\Omega} s_0 d \Omega = \int_{0}^{2 \pi } \int_{0}^{\pi } s_0 \sin \theta d \theta d \varphi = |a_{\ell m}|^2 + |b_{\ell m}|^2.
\end{equation}
Equation~\eqref{sigma_sca} shows that to determine $\sigma_{\rm{sca}}$,  $s_0$ needs to be measured in all directions. This measurement can be achieved using an integrating sphere (see Fig.~\ref{Fig_1}), something that is experimentally demanding. Even if we can experimentally measure $\sigma_{\rm{sca}}$ using an integrating sphere~\cite{kuznetsov2012magnetic}, distinguishing between the electric and magnetic amplitudes of the scattering coefficients in $\sigma_{\rm{sca}}$ is impossible, both are combined~\cite{luk2010fano, kuznetsov2012magnetic, garcia2011strong, luk2015optimum}. 
Our analytical findings, summarized in Eqs~\eqref{compact}-\eqref{vector}, provide a solution to this fundamental experimental limitation. We can now determine $|a_{\ell m}|^2$ and $|b_{\ell m}|^2$ separately from a measurement of the Stokes vector (see Fig.~\ref{Fig_1}b). This advancement allows us to differentiate between electric and magnetic resonances in objects that are well-described by a single multipolar order $\ell$ and total angular momentum $m$. 
Moreover, our Stokes polarimetry method allows us to capture optical anapoles. In a few words, anapoles are non-radiating sources whose signature is a dip in the scattering cross-section~\cite{miroshnichenko2015nonradiating, parker2020excitation}. In essence, when $|a_{\ell m}| = 0$ ($|b_{\ell m}| = 0$), an electric (magnetic) optical anapole emerges~\cite{wei2016excitation}.  When both $|a_{\ell m}| = |b_{\ell m}| =0$, an hybrid anapole arises~\cite{luk2017hybrid, canos2021theory, sanz2021multiple}.  Eqs~\eqref{compact}-\eqref{vector} indicate that one can unravel optical anapoles by a  measurement of the Stokes vector. Remarkably, we can also differentiate the nature of the optical anapole (electric, magnetic, or hybrid) upon this Stokes measurement.

\begin{figure}[t!]
    \centering
    \includegraphics[width=0.999\columnwidth]{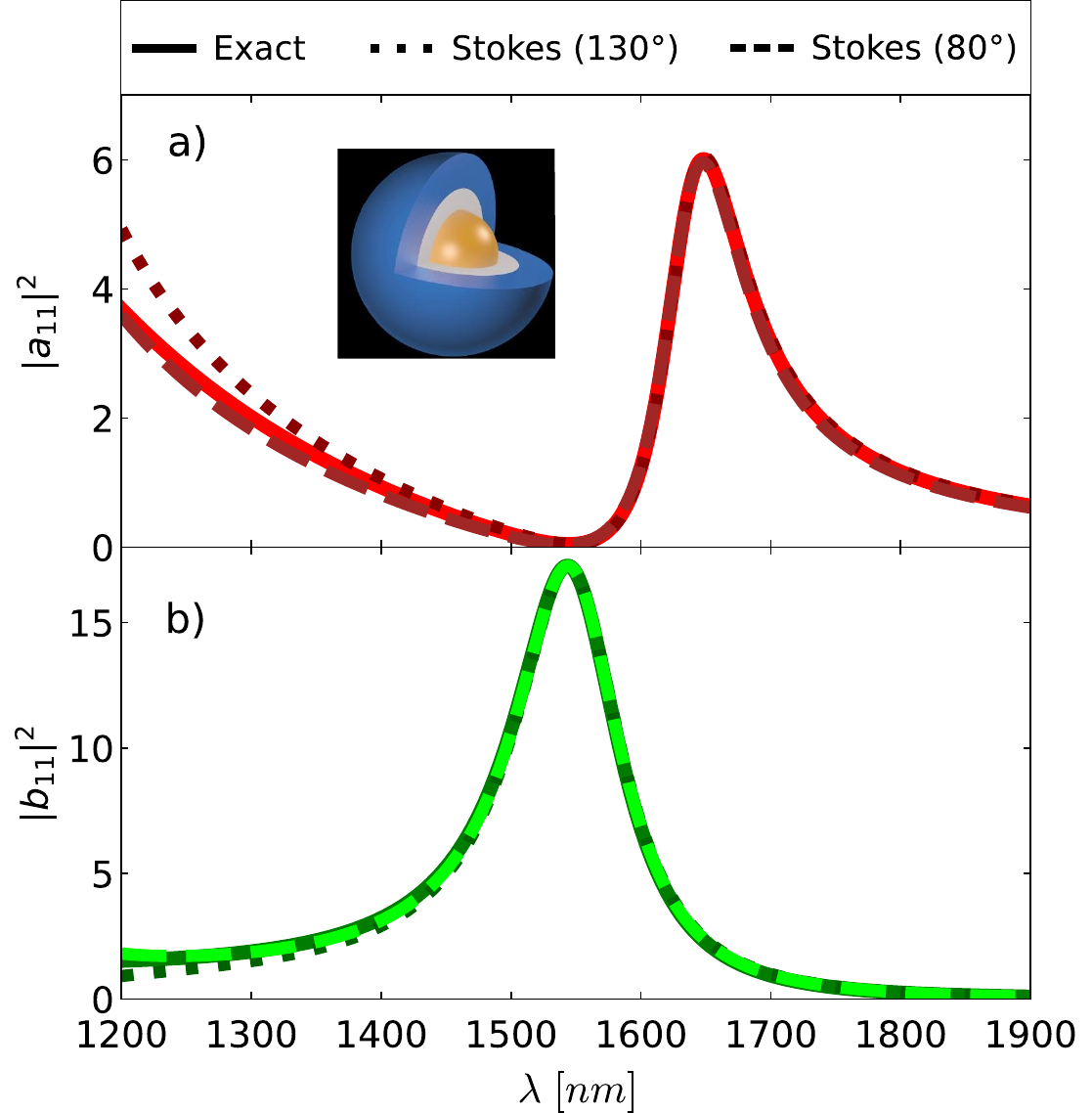}
    \caption{Quadratic combinations of the dipolar electric and magnetic scattering coefficients of an Au core-Ge shell nanoparticle embedded in air with an outer radius $b = 183$ nm and an inner radius $a = 63$ nm, respectively. The incident wavefield is a circularly polarized plane wave. These quadratic forms are calculated from Mie theory (solid) and using the Stokes polarimetry approach summarized in Table~\ref{T_1}. Two angles are chosen: $\theta = 130^\circ$ (dotted) and $\theta = 80^\circ$ (dashed). a) Electric amplitude $|a_{11}|^2$ depicted in red colors. b) Magnetic amplitude $|b_{11}|^2$ showed in green colors.}
    \label{Fig_2}
\end{figure}

\begin{figure}[t!]
    \centering
    \includegraphics[width=1\columnwidth]{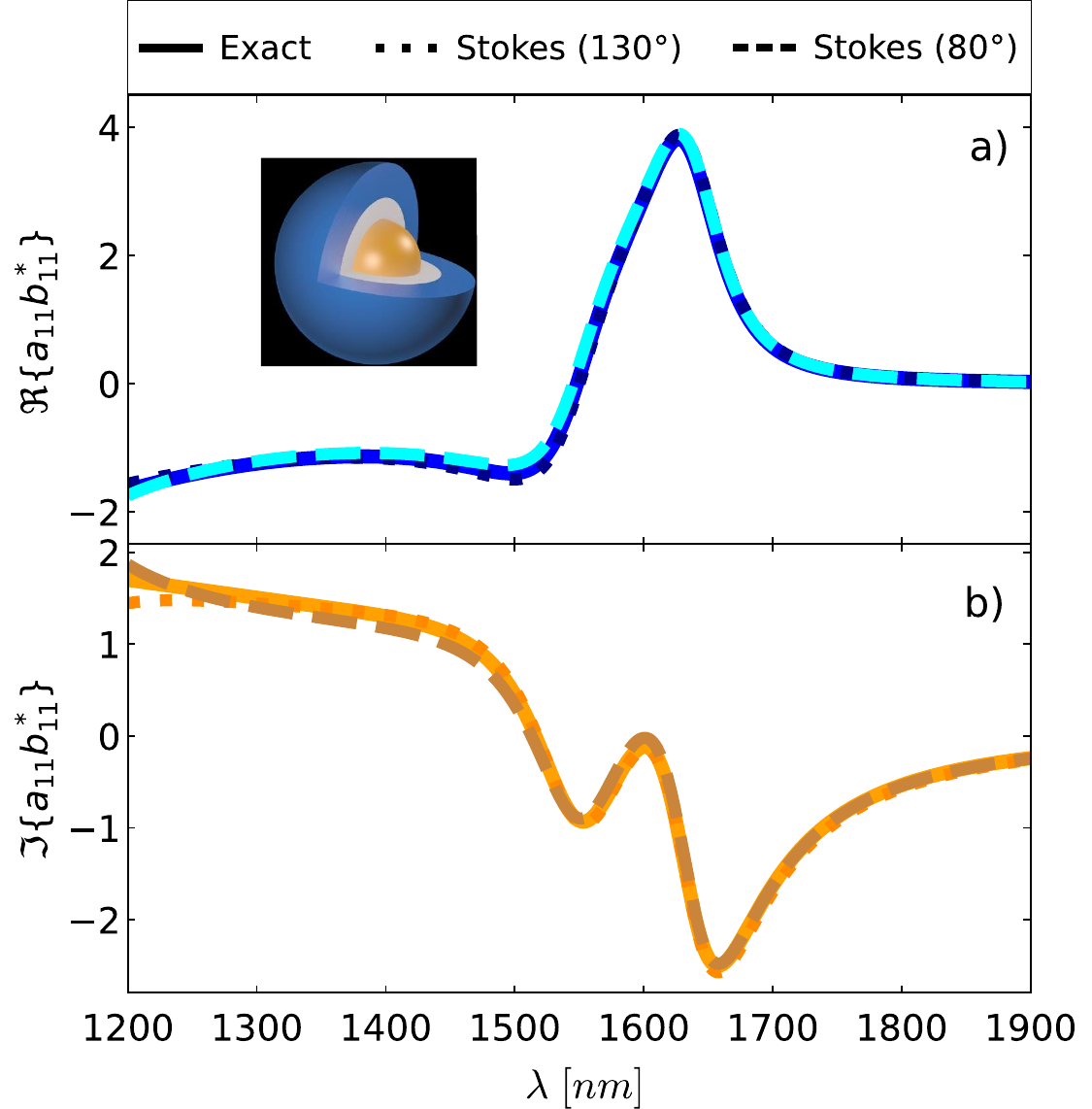}
    \caption{Quadratic combinations of the dipolar electric and magnetic scattering coefficients of an Au core-Ge shell nanoparticle embedded in air with an outer radius $b = 183$ nm and an inner radius $a = 63$ nm, respectively. The incident wavefield is a circularly polarized plane wave. These quadratic forms are calculated from Mie theory (solid) and using the Stokes polarimetry approach summarized in Table~\ref{T_1}. Two angles are chosen: $\theta = 130^\circ$ (dotted) and $\theta = 80^\circ$ (dashed). a) $\Re \{a_{11} b^*_{11} \}$ plotted in blue colors. b) $\Im \{a_{11} b^*_{11} \}$ depicted in orange colors. }
    \label{Fig_3}
\end{figure}

At this point, let us provide an illustrative example to show our Stokes polarimetry method in action. In particular, let us consider a Au core-Ge shell nanoparticle embedded in air with an inner radius $a = 63$ nm and an outer radius $b = 183$ nm, respectively. We anticipate that this object fulfills the following features when exposed to a plane wave:
\begin{itemize}

    \item At a certain wavelength, this object behaves as an ideal magnetic dipole~\cite{feng2017ideal}. In other words, at the magnetic dipolar resonance, the electric dipole vanishes.
\end{itemize}

In Fig.~\ref{Fig_2}a-b, we show the dipolar electric and magnetic amplitudes of this core-shell nanoparticle, given $|a_{11}|^2$ (see Fig.~\ref{Fig_2}a) and $|b_{11}|^2$ (see Fig.~\ref{Fig_2}b) . These dipolar amplitudes are determined using Mie theory (depicted by solid lines) and using our Stokes polarimetry approach evaluated at $\theta = 130^\circ$ (dotted lines) and $\theta = 80^\circ$ (dashed lines). Note that other scattering angles could have been selected. As depicted
in Fig.~\ref{Fig_2}a-b, the calculation of the electric and magnetic amplitudes from the
Stokes measurements shows an excellent agreement with the exact calculation in the broadband wavelength interval of $1400$ nm $<$ $\lambda$ $<$ $1900$ nm. As we have anticipated,  our Stokes polarimetry approach accurately captures the ideal magnetic dipole at $\lambda = 1540$ nm. Note that the results obtained from the Stokes polarimetry approach slightly deviate from each other (and from the exact result) at shorter wavelengths, specifically, $1200$ nm $<$ $\lambda$ $<$ $1400$ nm. This deviation occurs since, in this wavelength interval, the scattering cannot be fully described by $\ell = m =1$ due to the presence of the magnetic quadrupole. As a matter of fact, our Stokes polarimetry approach detects this non-negligible contribution of the magnetic quadrupole, serving as an explicit demonstration of the robustness of our method.  Note that numerical methods are not needed to infer the fidelity of our Stokes polarimetry approach. It is self-consistent. For a detailed explanation regarding the fidelity of our Stokes polarimetry approach, read Table~\ref{T_1}, in particular point 5.


\textbullet \; $\Re \{a_{\ell m} b^*_{\ell m} \}$ and $ \Im \{a_{\ell m} b^*_{\ell m} \}$. At this point, let us turn our attention to the interference terms of the quadrivector $\mathbf{D}_{\ell m}$.
These interference terms, namely, $\Re \{a_{\ell m} b^*_{\ell m} \}$ and $ \Im \{a_{\ell m} b^*_{\ell m} \}$, have not been as well-studied as the scattering cross-section in scattering theory. 
Fortunately, recent developments have shed light on these interference terms within the framework of the Generalized Lorentz Mie theory~\cite{gouesbet2011generalized}.
Briefly, the GLMT gives the exact solution of a spherical particle under general illumination conditions~\cite{gouesbet2011generalized}. 
Considering the GLMT,  we can write the interference terms of $\mathbf{D}_{\ell m}$ as 
\begin{align} \label{Mie_re}
\Re \{a_{\ell m} b^*_{\ell m} \} &= \Re \{g^{e}_{\ell m} {g^{m}_{\ell m}}^* \} \Re \{a_{\ell } b^*_{\ell} \} - \Im \{g^{e}_{\ell m} {g^{m}_{\ell m}}^* \} \Im \{a_{\ell } b^*_{\ell} \}, \\ \label{Mie_im}
\Im \{a_{\ell m} b^*_{\ell m} \} &= \Im \{g^{e}_{\ell m} {g^{m}_{\ell m}}^* \} \Re \{a_{\ell } b^*_{\ell} \} + \Re \{g^{e}_{\ell m} {g^{m}_{\ell m}}^* \} \Im \{a_{\ell } b^*_{\ell} \}.
\end{align}
Here, we have made use of $a_{\ell m} = - a_\ell g^{e}_{\ell m}$ and $b_{\ell m} = - b_\ell g^{m}_{\ell m}$, where $\{a_\ell, b_\ell \}$ are the electric and magnetic  Mie coefficients, respectively, and $\{g^{e}_{\ell m}, g^{m}_{\ell m}\}$ are the electric and magnetic coefficients characterizing the incident wavefield, respectively~\cite{olmos2023capturing}. 

We now reach notable results:
Eqs.~\eqref{Mie_re}-\eqref{Mie_im} show that one can  retrieve $\Re \{a_{\ell } b^*_{\ell} \}$ and $\Im \{a_{\ell } b^*_{\ell} \}$ separately upon a Stokes vector measurement. Let us show this by manipulating the helicity of the incident wavefield, with eigenvalues  $\sigma  = \pm 1$. 
First, we note that a wavefield carrying well-defined helicity $\sigma = +1$ satisfies $g^{e}_{\ell m} = -i g^{m}_{\ell m}$~\cite{olmos2023capturing}, and hence, yields $\Re \{g^{e}_{\ell m} {g^{m}_{\ell m}}^* \} = 0$. In this setting, Eqs.~\eqref{Mie_re}-\eqref{Mie_im} are simplified to 
\begin{align} \label{Mie_re_hel}
\frac{\Re \{a_{\ell m} b^*_{\ell m} \}}{|g^{e}_{\ell m}|^2} =  \Im \{a_{\ell } b^*_{\ell} \},  &&
\frac{\Im \{a_{\ell m} b^*_{\ell m} \}}{|g^{e}_{\ell m}|^2} = -   \Re \{a_{\ell } b^*_{\ell} \}.
\end{align}
Equation~\eqref{Mie_re_hel} shows that the inference terms between the electric and magnetic Mie coefficients, namely, $ \Re \{a_{\ell } b^*_{\ell} \}$ and $ \Im \{a_{\ell } b^*_{\ell} \} $,  can be separately determined from a  measurement of the Stokes vector. 
Indeed, in Fig.~\ref{Fig_3}, we show these interference terms, namely,  $ \Re \{a_{11} b^*_{11} \} = |g^e_{11}|^2 \Im \{a_{1 } b^*_{1} \} $ (see Fig.~\ref{Fig_3}a)  and $ \Im \{a_{11} b^*_{11} \} = -|g^e_{11}|^2 \Re \{a_{\ell } b^*_{\ell} \}$ (see Fig.~\ref{Fig_3}b),  obtained from Mie theory and using our Stokes polarimetry approach evaluated at the previous scattering angles, namely, $\theta = 130^\circ$ and $\theta = 80^\circ$. From Fig.~\ref{Fig_3}, we can note that there is a remarkable agreement between both calculations in the wavelength interval of $1400$ nm $<$ $\lambda$ $<$ $1900$ nm, pointing out that our Stokes polarimetry approach, summarized in Eqs~\eqref{compact}-\eqref{vector}, is suitable to retrieve the interference terms.
To the best of our knowledge, there is currently no alternative method to experimentally measure these interference terms from a single measurement of the Stokes vector.
Having noted this, these interference terms have recently emerged as key quantities in various branches of Nanophotonics.
For instance, the interference term $\Re \{a_{\ell} b^*_{\ell} \}$ has shown to be of utmost significance in the preservation of helicity~\cite{hanifeh2020helicity}, Kerker conditions~\cite{geffrin2012magnetic,  paniagua2011metallo, olmos2020unveiling, olmos2020kerker, olmos2020optimal}, surface-enhanced circular dichroism enhancements~\cite{garcia2013surface}, light transport phenomena~\cite{gomez2012negative}, and optical forces~\cite{nieto2011angle, nieto2010optical, gomez2012electric, xu2019azimuthal}. 
In stark contrast, $\Im \{a_{\ell} b^*_{\ell} \}$ has remained relatively unexplored until recently, primarily appearing in the context of spinless optical mirages~\cite{olmos2023optical} and recoiling optical forces~\cite{bekshaev2015transverse, xu2019azimuthal}. 

{\it{Conclusions.}}--- We have demonstrated that a measurement of the Stokes vector unlocks key magnitudes at the core of Nanophotonics. These magnitudes are constructed from the quadrivector $ \mathbf{D}_{\ell m}  = \left[|a_{\ell m}|^2, |b_{\ell m}|^2, \Re \{ a_{\ell m} b^*_{\ell m}  \}, \Im \{ a_{\ell m} b^*_{\ell m}  \} \right]$, captured using our Stokes polarimetry approach.
We have shown that the determination of $\mathbf{D}_{\ell m}$ grants access to:
\begin{itemize}
    \item The separate detection of the amplitudes $|a_{\ell m}|^2$ and $|b_{\ell m}|^2$. Remarkably, this distinction is unattainable if measuring the total energy of the scattered field via an integrating sphere. 
    \item The detection of the interference terms between the electric and magnetic Mie coefficients from the same Stokes vector measurement. We have disentangled these interference terms by manipulating the incident helicty of the wavefield.
\end{itemize}

As a final remark, we have demonstrated that our Stokes polarimetry approach is robust and can be fully trusted upon two measurements of the Stokes vector. Thereby, our findings hold significant potential for a wide variety of Nanophotonics branches as they greatly facilitate the experimental characterization of objects in optical laboratories.

\section*{Acknowledgements}
J.O-T. acknowledges Adrian Juan-Delgado and Dr. Cristina Sánz-Fernández for useful comments.
J.O-T  acknowledges support from the Juan de la Cierva fellowship No. FJC2021-047090-I of  MCIN/AEI/10.13039/501100011033 and NextGenerationEU/PRTR and acknowledges financial support from the Spanish
Ministry of Science and Innovation (MCIN), AEI and
FEDER (UE) through project PID2022-137569NB-C43.

\section*{Disclosures}
The authors declare no conflict of interest.

\clearpage

\section*{References}

\bibliography{Bib_tesis} 

\clearpage

\appendix

\section{The Scattered Electromagnetic field in the Far-field} \label{A_1}

In this Appendix,  we determine the complex amplitudes of the transversal components $E_\theta $ and $E_\varphi$ presented in the main text (see Eqs.~\eqref{key_1}-\eqref{key_2}).  Let us start by writing the scattered electromagnetic field $\E(k \r)$ in  terms of electric and magnetic multipoles~\cite{olmos2023capturing},
\begin{equation} \label{E_sca}
\frac{\E(k \r)}{E_0} = \sum_{\ell m} \left[ a_{\ell m}\boldsymbol{N}_{\ell m}(k \r) +  b_{\ell m}\boldsymbol{M}_{\ell m}(k \r) \right].
\end{equation}
Here  $\boldsymbol{M}_{\ell m}(k \r) =h^{(1)}_\ell(kr)\boldsymbol{X}_{\ell m}(\r) $ and $k \boldsymbol{N}_{\ell m}(k \r) = i \nabla \times \boldsymbol{M}_{\ell m}( k \r)$ are Hansel multipoles~\cite{olmos2023capturing}, $\boldsymbol{X}_{\ell m}(\hat{\r})  = {\mathbf{L}  Y_{\ell m}(\theta, \varphi)}/{\sqrt{\ell(\ell +1 )}}$ are vector spherical harmonics,  $h^{(1)}_\ell (kr)$ are the spherical Hankel function of the first kind, $k$ is the radiation wavelength,  $r = |\r|$ denotes the observation point,  $\theta$ and $\varphi$ are the scattering and azimuthal angle, respectively. In this framework,  $\mathbf{L}= -i \r \times \nabla$ is the total angular momentum operator and $Y_{\ell m}(\theta, \varphi)$ are spherical harmonics  defined as in Ref.~\cite{jackson1999electrodynamics}
\begin{equation} \label{Y_lm}
    Y_{\ell m} (\theta, \varphi) = \sqrt{\frac{2 \ell +1}{4 \pi }\frac{(\ell - m)!}{(\ell+m)!}}e^{im \varphi}   P^{m}_{\ell}(\cos \theta),
\end{equation}
where $P^{m}_{\ell}(\cos \theta)$ are the associated Legendre Polynomials~\cite{jackson1999electrodynamics}. 
Moreover,  $a_{lm}$ and $b_{lm}$ stand for the (dimensionless)  electric and magnetic scattering coefficients, respectively, $\ell$ and $m$ being the multipolar order and total angular momentum of the scattered electromagnetic field introduced in Eq.~\eqref{E_sca}, respectively.

Now, let's calculate Eq.~\eqref{E_sca} in the far-field limit, namely, when $ kr \rightarrow \infty$. After algebra, we arrive from Eq.~\eqref{E_sca} to  
\begin{equation} \label{E_far}
\frac{\E(k \r)}{E_0} = \frac{ i e^{ikr}}{kr} \left[  (-i)^{\ell} \left[ a_{\ell m}(\hat{\r} \times \boldsymbol{X}_{\ell m}(\hat{\r})) - b_{\ell m}\boldsymbol{X}_{\ell m}(\hat{\r})\right] \right], 
\end{equation}
where we have made use of the following relations~\cite{jackson1999electrodynamics}
\be 
\lim_{kr \rightarrow \infty} \boldsymbol{N}_{\ell m}(k \r) &=& i \frac{e^{ikr}}{kr} (-i)^\ell  (\hat{\r} \times \boldsymbol{X}_{\ell m}(\hat{\r})), \\
\lim_{kr \rightarrow \infty} \boldsymbol{M}_{\ell m}(k \r) &=& \frac{e^{ikr}}{kr} (-i)^{\ell+1}  \boldsymbol{X}_{\ell m}(\hat{\r}).
\ee 
At this point, let us express the vector spherical harmonics $\boldsymbol{X}_{\ell m}(\hat{\r})$ in spherical coordinates~\cite{jackson1999electrodynamics}. 
The  total angular momentum operator $\mathbf{L}$ reads in spherical coordinates as  
\begin{equation} \label{L}
\mathbf{L} =     -i \left[- \hat{\mathbf{e}}_{\theta} \frac{1}{\sin \theta} \frac{\partial}{\partial_{\varphi}}    + \hat{\mathbf{e}}_{\varphi} \frac{\partial}{\partial_{\theta}} \right].
\end{equation}
Therefore, we can write $\boldsymbol{X}_{\ell m}(\hat{\r})$ in spherical coordinates as 
\begin{equation} \label{X}
\boldsymbol{X}_{\ell m}(\hat{\r}) = \frac{-i}{\sqrt{\ell ( \ell + 1) }} \left[- \hat{\mathbf{e}}_{\theta} \frac{1}{\sin \theta} \frac{\partial}{\partial_{\varphi}}    + \hat{\mathbf{e}}_{\varphi} \frac{\partial}{\partial_{\theta}} \right] Y_{\ell m} (\theta, \varphi).
\end{equation}
Now, by taking into account Eq.~\eqref{Y_lm}, we can write   
\begin{align} 
\boldsymbol{X}_{\ell m}(\hat{\r}) &=& C_{\ell  m} (\varphi)\left(- i m \pi_{\ell m} (\theta ) \hat{\mathbf{e}}_{\theta} + \tau_{\ell m}(\theta) \hat{\mathbf{e}}_{\varphi}  \right), \\ \label{rotX}
\hat{\r} \times \boldsymbol{X}_{\ell m}(\hat{\r}) &=& -C_{\ell  m} (\varphi) \left(\tau_{\ell m}(\theta) \hat{\mathbf{e}}_{\theta} +i m \pi_{\ell m}(\theta )\hat{\mathbf{e}}_{\varphi}  \right),
\end{align}
where we have defined 
\begin{align}
\pi_{\ell m}(\theta) =  \frac{P^{m}_{\ell }(\cos \theta)}{\sin \theta}, &&    \tau_{\ell m}(\theta)  = \frac{d P^{m}_{\ell }(\cos \theta)}{d{\theta}},
\end{align}
and
\begin{equation}
C_{\ell  m} (\varphi)= \frac{-i}{\sqrt{\ell ( \ell + 1) }} \sqrt{\frac{2 \ell +1}{4\pi}\frac{(\ell - m)!}{(\ell+m)!}}e^{im \varphi}.
\end{equation} 
At this point, let us insert Eqs.~\eqref{X}-\eqref{rotX} into Eq~\eqref{E_far}. After some algebraic manipulation, it can be shown that 
\begin{equation}
\lim_{kr \rightarrow \infty} {\E (k \r)}=  \left[ E_\theta \hat{\mathbf{e}}_{{{\theta}}} + E_\varphi \hat{\mathbf{e}}_{{{\varphi}}}\right], 
\end{equation}
where 
\be 
E_\theta = {E_0} \sum_{\ell m} \bar{C}_{\ell  m} (kr, \varphi) \left[  a_{\ell m} \tau_{\ell m} ({\theta}) - im b_{\ell m} \pi_{\ell m} ({\theta}) \right],  \\
E_\varphi = {E_0} \sum_{\ell m} \bar{C}_{\ell  m} (kr, \varphi) \left[ im  a_{\ell m} \pi_{\ell m} ({\theta}) + b_{\ell m} \tau_{\ell m} ({\theta}) \right],
\ee
where $\bar{C}_{\ell m}(kr, \varphi) = (-i)^{\ell + 1}  \cfrac{e^{ikr}}{kr} C_{\ell m}(\varphi)$.

\section{The $U_{11}$ matrix} \label{A_2}
In this Appendix, we calculate the $U_{\ell m}$ matrix for $\ell = m =1$. Now, by inspecting Eq.~\eqref{matrix_change} we need the Associated Legendre Polynomials $P_{\ell m}(\cos \theta)$. Hereafter, we follow the notation of Jackson's book in its third edition~\cite{jackson1999electrodynamics}. When setting $\ell = m = 1$, we get $P_{11}(\cos \theta) = -\sin \theta$. Then, we insert this into Eq.~\eqref{chi_tau} of the main text to obtain, $\tau_{11} = -1$ and $\pi_{11} = - \cos \theta$. Then, we insert these values in Eqs. (14-16) yielding
\begin{align} \label{B1}
\gamma_{11} = 1 + \cos^2 \theta,     && \eta_{11} =\cos \theta, && \nu_{11} = -\sin^2 \theta.
\end{align}
At this point, only a step remains to be done to calculate $U_{11}$ matrix: we need to compute  ${A}_{11} = 2\nu_{11}^2 |E_0|^2 |C_{11}|^2$. To achieve this goal, we must compute $|C_{11}|^2$. This expression can be found in Eq.~\eqref{constant}. Now, by setting $\ell = m = 1$, we get
\begin{equation} \label{B2}
|C_{11}|^2 = \frac{3 }{16 \pi}\frac{1}{(kr)^2}
\end{equation}
At this point, we have all the ingredients to calculate the  $U_{11}$ matrix. Taking into account Eqs.~\eqref{B1}-\eqref{B2}, we arrive to
\begin{align} \label{matrix_change_1}
U_{11} = \frac{1}{A_{11}}   
    \begin{pmatrix}
    1 + \cos^2 \theta & -\sin^2 \theta & 0 & 2 \cos \theta \\
    1 + \cos^2 \theta & \sin^2 \theta& 0 & 2 \cos \theta \\
       0 & 0 & \sin^2 \theta & 0\\
    2 \cos \theta & 0 & 0 &  1 + \cos^2 \theta
    \end{pmatrix},
\end{align}
where $A_{11} =  \cfrac{3|E_0|^2 \sin^4 \theta}{8 \pi (kr)^2}$.

\clearpage

\end{document}